# Predicting Process Name from Network Data

**Justin Allen, David Knapp, Kristine Monteith**
Lawrence Livermore National Laboratory
allen119@llnl.gov, knapp2@llnl.gov, monteith3@llnl.gov

## Abstract

The ability to identify applications based on the network data they generate could be a valuable tool for cyber defense. We report on a machine learning technique capable of using netflow-like features to predict the application that generated the traffic. In our experiments, we used ground-truth labels obtained from host-based sensors deployed in a large enterprise environment; we applied random forests and multilayer perceptrons to the tasks of browser vs. non-browser identification, browser fingerprinting, and process name prediction. For each of these tasks, we demonstrate how machine learning models can achieve high classification accuracy using only netflow-like features as the basis for classification.

## 1 Introduction

Cyber security analysts benefit from having as much information as possible for use in defending their networks and systems. With current technologies, many system issues can only be detected used host-based sensors, but because of the resource requirements for these types of sensors, data collected by network-based sensors may be more readily available. Accurate models of processes based solely on the network traffic they generate could provide system administrators with greater insight into the status of individual machines on their network without a host-based sensor. These models could be useful in automatic resource provisioning for specific processes to improve quality-of-service [Stewart *et al.*, 2005]. They could provide information about the operating systems and patch levels of systems on a network. An understanding of the applications that are active on the individual hosts of a network could also potentially allow for detection of malicious activity [Carl *et al.*, 2006; Mitropoulos *et al.*, 2017].

Many modeling efforts have focused on the identification of broad application types, not individual applications. And such models often rely on hard-coded rules, using ports or other simple features to identify the type of application responsible for a given network signature. Over time, changes in traffic patterns can drastically reduce the efficacy of hard-coded models. The use of dynamic ports can render direct port lookups useless. The increasing ubiquity of encryption often prevents the explicit use of packet content for process identification. In addition, even effective hard-coded models often require manual maintenance in response to changes in process behavior.

In contrast, we demonstrate how more robust machine learning models can be used to identify not only application type, but the specific process that generated the traffic. We employed a host-based collection from nearly 1200 distinct machines, with network traffic data collected on over 1400 unique processes, to these types of classification tasks. Our results show that netflow-like features can be used to differentiate between browser and non-browser traffic. For browser traffic, our models are able to identify the specific browser type with reasonably high accuracy. For non-browser traffic, our models can effectively predict the traffic-generating host process.

## 2 Literature Review

The concept of applying machine-learning techniques to the task of application identification is not new; initial work in this area focused on clustering general traffic type based on features such as packet length statistics, byte counts, and interarrival times. McGregor [2004] used an expectation-maximization algorithm to cluster traffic. Zander [2005] applied the AutoClass clustering algorithm. Bernaille [2006] employed *k*-means clustering.

Researchers have also applied supervised learning strategies to the task of traffic classification. Roughan [2004] proposed a technique using nearest neighbors and linear and quadratic discriminant analysis to predict up to seven classes of traffic (HTTP, FTP, etc.). Nguyen [2006] used a Naive Bayes model trained on a sliding window of netflow data to classify traffic. Park [2006] compared the efficacies of the Naive Bayes classifier with kernel estimation and decision trees. Auld [2007] used Bayesian neural networks on a larger set of features, including flow metrics, packet interarrival times, effective bandwidth, and size of TCP/IP control fields. Sun [2010] collected host-based data using a distributed host-based traffic collection platform and labeled data as 'Web,' 'P2P,' and 'Other;' probabilistic neural networks, support-vector machines, and radial-basis-function neural networks were used to model the traffic.

These models were trained on network data grouped by traffic type (P2P, WWW, FTP, HTTP, etc.) instead by individual traffic-generating application. Attempts to identify individual applications seem to be less common, likely due to the lack of available training sets. As an example of this type of task, Lin [2009] showed that processes tend to have a unique packet size distribution and port association, and that a small number of specific applications could be classified based on their internet traffic. Aceto [2018] used various machine learning methods to predict applications from encrypted mobile traffic. Liu [2016] used the C4.5 algorithm and packet sizes of encrypted traffic as features for the task of browser identification.

## 3 Experimental Set-Up

### 3.1 Host-Based Sensor Data

The data set for our experiments consisted of aggregate packet flow statistics generated from host-based sensors deployed on nearly 1200 unique hosts running a Windows operating system in an enterprise environment. The sensors monitor generated network traffic every ten seconds and associate it back with the process that generated it. A list of aggregate statistics is shown in Table 1.

| | |
|---|---|
| Protocol (TCP or UDP) | Ratio of TCP to UDP bytes sent |
| | Ratio of TCP to UDP bytes received |
| Total bytes sent | Ratio of TCP to UDP packets sent |
| Average packet size | Ratio of TCP to UDP packets received |
| Number of TCP bytes sent | |
| Number of TCP bytes received | |
| Number of TCP packets sent | |
| Number of TCP packets received | Total Events |
| Average TCP packet size | Count of TCP Accept events |
| Number of TCP bytes copied | Count of TCP Connect events |
| Number of TCP packets copied | Count of TCP Reconnect events |
| | Count of TCP Disconnect events |
| Number of UDP bytes sent | Count of TCP Receive events |
| Number of UDP bytes received | Count of TCP Retransmit events |
| Number of UDP packets sent | |
| Number of UDP packets received | |
| Average UDP packet size | |

Table 1: Aggregate packet flow statistics uses as features in classification experiments.

The full collect consisted of packet flow statistics from 565 million instances of process execution. However, a few process types (e.g., svchost.exe) accounted for a majority of the data. To generate a less skewed data set, we sampled up to 50,000 examples for each of the 1407 unique process names in the collect. Our subsampled data set consisted of 8.8 million data points.

### 3.2 Machine Learning Models

Random forests and multilayer perceptrons were used for modeling in our classification experiments.

Random forests were implemented using the scikit-learn RandomForestClassifier module, using the parameters *n_estimators*=100 and *max_depth*=15.

Multilayer perceptrons were implemented using Keras over Tensorflow. For these models, each network consisted of an input layer, three hidden layers, and an output layer, with the size of each layer decreasing logarithmically from the input to the output size. ReLU activation was used on all layers except for the last, which had a softmax activation. The networks employed a cross-entropy loss function and the Nesterov-Adam optimizer. Batch normalization was used to reduce overfitting.

Several of the calculated network statistics tended to have large raw values that would overwhelm the multilayer perceptrons. Consequently, a histogram was constructed of typical values for each feature; models were training based on bin numbers instead of raw values.

## 4 Results

Our modeling experiments included differentiating between browser and non-browser traffic, fingerprinting the browser responsible for the network traffic, and predicting the name of the traffic-generating host-process.

Information about the number of labels and data set size for each experiment can be found in Tables 2 and 3. These tables also report overall classification accuracy, average precision, and average recall. For each set of experiments, models were trained on 80% of the data and evaluated on 20%.

### 4.1 Browser vs. Non-Browser

For our first classification task, we trained models to differentiate between browser and non-browser traffic. Activity from all browsers in our data set (Firefox, Chrome, Internet Explorer, and Microsoft Edge) was labeled as positive examples, while all other traffic was labeled as negative examples. Figure 1 shows the results of these classification experiments. Overall classification accuracy for both models was over 88%.

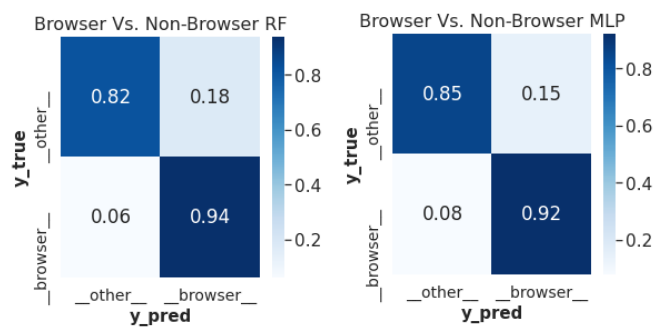

Figure 1: Confusion matrix for the browser vs. non-browser classification task using random forests (left) and multilayer perceptrons (right).

## 4.2 Browser Fingerprinting

The next set of classification tasks involved fingerprinting the particular browser that generated network traffic. Figures 2 and 3 show the ability of random forests and multilayer perceptrons, respectively, to differentiate between traffic from Firefox, Chrome, Internet Explorer, and Microsoft Edge.

Accuracies were slightly lower on this task compared to the simple browser *vs.* non-browser classification. Overall classification accuracy was 74% for the random forest and 72% for the multilayer perceptron. Not surprisingly, classifiers had some difficulty differentiating between Internet Explorer and Microsoft Edge traffic. There was also some confusion between Firefox and Chrome traffic.

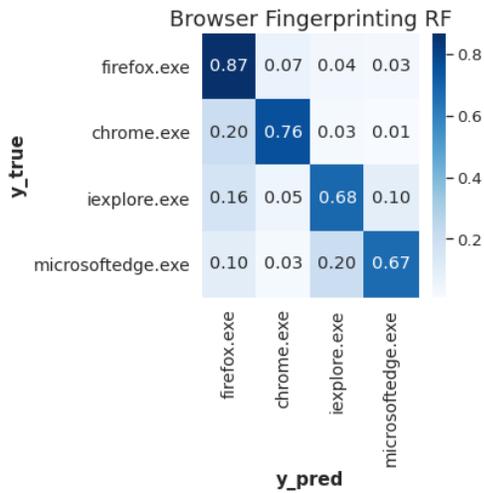

Figure 2: Confusion matrix for the browser fingerprinting task using random forests.

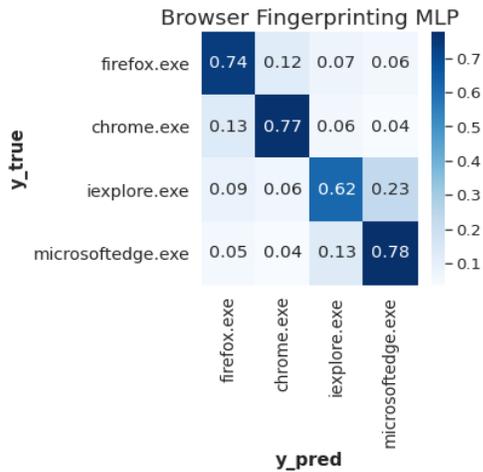

Figure 3: Confusion matrix for the browser fingerprinting task using multilayer perceptrons.

## 4.3 Browser Fingerprinting and Non-Browser Identification

The two previous tasks were combined in a third task, in which models were trained both to differentiate between traffic from different browsers and to separate it from non-browser traffic. Results are shown in Figures 4 and 5. In this case, both models achieved an overall classification accuracy of 71%.

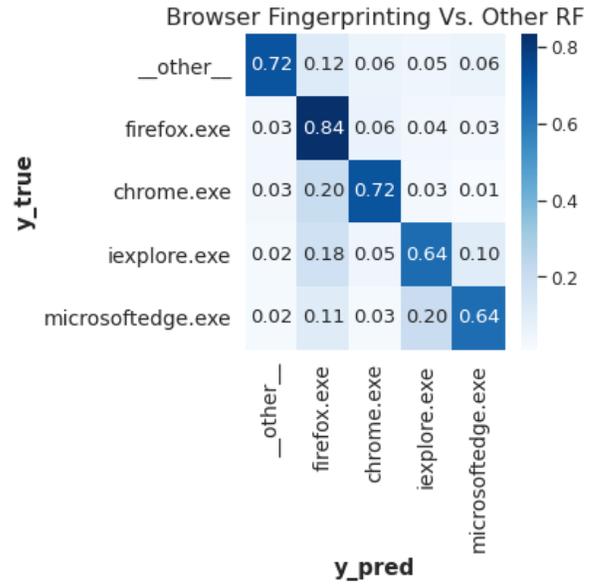

Figure 4: Confusion matrix for the browser fingerprinting and non-browser identification task using random forests.

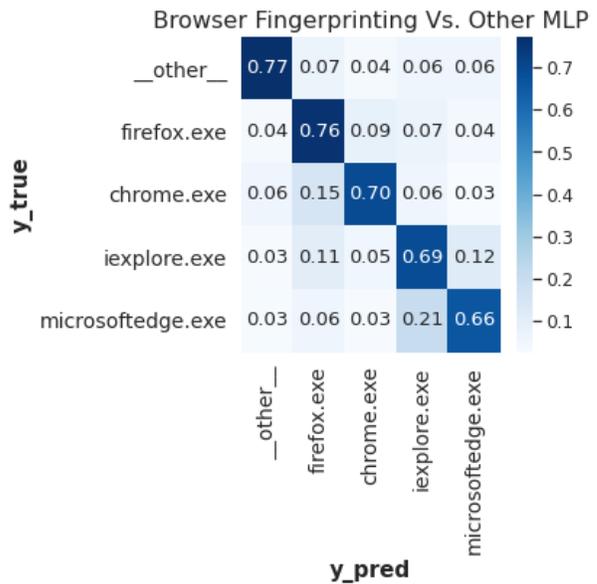

Figure 5: Confusion matrix for the browser fingerprinting and non-browser identification task using multilayer perceptrons.

## 4.4 Process Name Prediction

For our final set of classification tasks, we trained models to differentiate between hundreds of traffic-generating processes in our data set. Any process name with at least 300 samples of network traffic was included in the data set. As with the previous experiments, the number of samples was capped at 50,000 for any given process. 484 unique process names were considered, with over 8 million samples used in training and testing.

Our multilayer perceptrons performed better on this classification task than our random forest models; they achieved an overall classification accuracy of almost 70%. Classification accuracy of the random forests was 63%. The confusion matrix for the multilayer perceptron experiment is shown in Figure 6. The noticeable mass on the diagonal suggests that models can often predict process names with as few as 300 training examples and hundreds of competing processes.

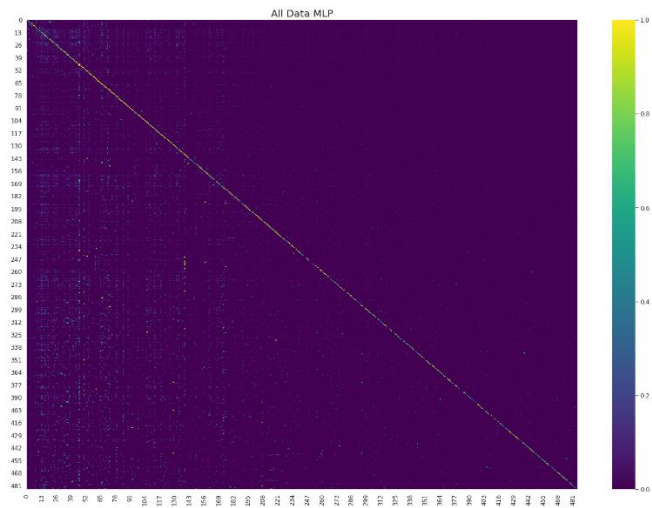

Figure 6: Confusion matrix for the process name prediction task using multilayer perceptrons.

We repeated these experiments using just the top $N$ processes for the following values of $N$: 5, 10, 50, 100, 300, 500, and 1000. In each case, models were trained to differentiate between the top $N$ processes, with remaining processes grouped into an "Other" category.

Figure 7 illustrates how overall classification accuracy, average precision, and average recall degraded with an increase in the number of process names considered. Note that process names beyond the top 484 most common had less than 300 samples in these data sets; it is not surprising that average precision and recall drop considerably when these less common process names are included.

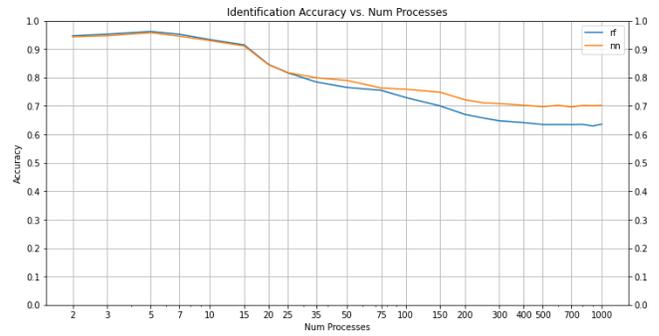
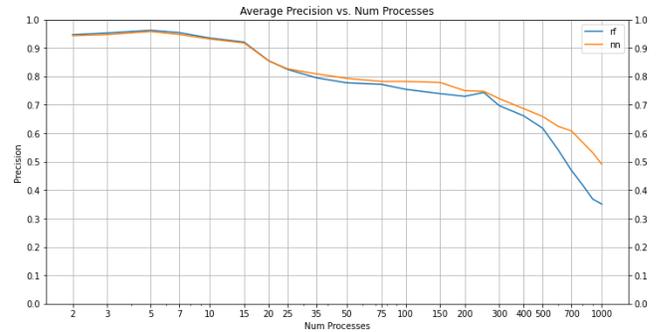
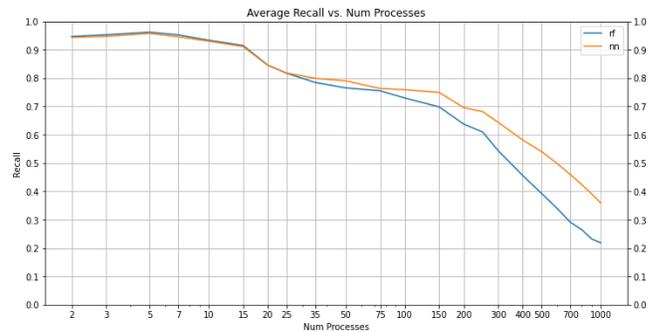

Figure 7: Average classification accuracy, precision, and recall degraded with an increase in the number of process names considered.

## 5 Discussion

We have demonstrated the efficacy of two common machine learning models for application-specific internet traffic classification. Using only the protocol and aggregate statistics of network traffic, our models were able to differentiate between browser and non-browser traffic with accuracies as high as 88%. Models could identify the browser that generated network traffic with overall classification accuracies in the mid-70%. They were also able to effectively differentiate between hundreds of unique processes with nearly 70% overall classification accuracy. In some cases, processes with as few as 300 samples of network traffic were correctly labeled.

These experiments establish a proof-of-concept for the identification of host-based processes based on the network traffic that they generate. Further experiments are necessary to determine how well these results generalize, and how effectively this information can be used for specific cyber defense tasks.

Degree of generalization is difficult to assess; privacy and security concerns generally prevent sharing data from host-based sensors between organizations. (Note that these experiments are only possible because of the generosity of the data owners in making it available in a limited fashion for research purposes.) However, if further experiments confirm that models can generalize from one environment to another, then such models could be trained in a highly instrumented environment and deployed in a less instrumented one.

Further research is also required to determine the suitability of these types of models for specific cyber defense tasks. For example, future experiments could involve determining whether network traffic changes depending on the patch level of a given system, or whether a portfolio of processes for a specific threat could be detected by the network traffic they generate.

| | Number of Samples | Number of Labels | Model | Classification Accuracy | Average Precision | Average Recall |
|---|---|---|---|---|---|---|
| Browser vs. Non-Browser | 800,000 | 2 | RF | 88.04% | 88.57% | 88.04% |
| | | | MLP | 88.73% | 88.91% | 88.73% |
| Browser Prediction | 400,000 | 4 | RF | 74.49% | 75.89% | 74.49% |
| | | | MLP | 72.71% | 72.70% | 72.71% |
| Browser Prediction + Non-Browser | 500,000 | 5 | RF | 71.29% | 73.19% | 71.29% |
| | | | MLP | 71.72% | 72.29% | 71.72% |

Table 2: Results of modeling browser vs. non-broswer traffic and predicting the traffic-generating browser.

| | Number of Samples | Number of Labels | Model | Classification Accuracy | Average Precision | Average Recall |
|---|---|---|---|---|---|---|
| All Processes | 8,262,317 | 484 | RF | 63.58% | 55.59% | 37.87% |
| | | | MLP | 69.84% | 66.26% | 54.97% |
| Top 5 | 250,000 | 5 | RF | 96.23% | 96.26% | 96.23% |
| | | | MLP | 95.78% | 95.81% | 95.78% |
| Top 10 | 500,000 | 10 | RF | 93.38% | 93.53% | 93.38% |
| | | | MLP | 93.03% | 93.18% | 93.03% |
| Top 50 | 2,250,000 | 50 | RF | 76.55% | 77.75% | 76.55% |
| | | | MLP | 79.01% | 79.35% | 79.01% |
| Top 100 | 4,600,000 | 100 | RF | 72.99% | 75.46% | 72.99% |
| | | | MLP | 75.90% | 78.25% | 75.90% |
| Top 300 | 8,090,040 | 300 | RF | 64.78% | 69.73% | 54.37% |
| | | | MLP | 70.84% | 72.20% | 64.35% |
| Top 500 | 8,267,174 | 500 | RF | 63.46% | 61.76% | 39.22% |
| | | | MLP | 69.84% | 65.89% | 54.02% |
| Top 1000 | 8,304,209 | 1000 | RF | 63.61% | 35.07% | 21.83% |
| | | | MLP | 70.20% | 49.10% | 35.92% |

Table 3: Results of models predicting process name.